# Prediction of Probabilistic Transient Stability Using Support Vector Machine

Umair Shahzad

*Abstract*—Transient stability assessment is an integral part of dynamic security assessment of power systems. Traditional methods of transient stability assessment, such as time domain simulation approach and direct methods, are appropriate for offline studies and thus, cannot be applied for online transient stability prediction, which is a major requirement in modern power systems. This motivated the requirement to apply an artificial intelligence-based approach. In this regard, supervised machine learning is beneficial for predicting transient stability status, in the presence of uncertainties. Therefore, this paper examines the application of a binary support vector machine-based supervised machine learning, for predicting the transient stability status of a power system, considering uncertainties of various factors, such as load, faulted line, fault type, fault location and fault clearing time. The support vector machine is trained using a Gaussian Radial Basis function kernel and its hyperparameters are optimized using Bayesian optimization. Results obtained for the IEEE 14-bus test system demonstrated that the proposed method offers a fast technique for probabilistic transient stability status prediction, with an excellent accuracy. DIgSILENT PowerFactory and MATLAB was utilized for transient stability time-domain simulations (for obtaining training data for support vector machine) and for applying support vector machine, respectively.

*Index Terms*—Artificial Intelligence (AI), dynamic security assessment (DSA), machine learning, probabilistic transient stability, support vector machine (SVM).

## I. Introduction

THE reliability of a power system is defined as the ability of the power system to provide electric energy to consumers on a continuous basis and with acceptable service quality, for both planned and unplanned outages [1]. One of the main requirements to maintain the reliability of the power is to continually operate the synchronous generators and with satisfactory capacity to satisfy the load. In the domain of power system stability, transient stability is the ability of the synchronous machines to remain in synchronism during few seconds (usually 5-10 seconds), after a large disturbance, such as a short circuit fault, occurs [2]. In addition to fault type, fault location, system inertia, and system load, fault clearing time (FCT) and critical clearing time (CCT) are significant parameters in assessing transient stability [3]. FCT is the time at which fault is cleared after fault occurrence, whereas, CCT is the maximum FCT after which the system becomes transiently unstable [3]. Transient stability is an integral component of dynamic security assessment (DSA). DSA deals with evaluation of transient performance of the system after the occurrence of contingency [2]. To evaluate the transient stability status, time-domain simulation approach is generally used to solve a set of nonlinear differential-algebraic equations, which represent the dynamics of the network [4]. Although, the time-domain approach is the most accurate method and usually yields promising results, it is time consuming, as it must traverse a set of differential-algebraic equations, which can be computationally intensive, especially for large-scale systems [5]. The transient energy function (TEF) method [6]-[7], and the extended equal area criterion (EEAC) [8] have also been applied to transient stability evaluation; however, these approaches have some restrictions regarding modeling, and they require many computations to evaluate an index for transient stability status [9]. Moreover, these methods are not appropriate for online transient stability prediction [10].

Conventionally, deterministic methods, employing worst-case scenarios (seasonal peak load, three-phase fault, etc.), have been used to evaluate transient stability [11]-[13]. These approaches are very conservative and ignore the probability of various input parameters linked with transient stability, such as load, fault type, fault location, etc. With the continuous integration of renewable generation, the increasing prevalence of competitive electricity market, and the rising uncertainties in the power system, these conventional methods are becoming obsolete and unsuitable. Compared with the deterministic assessment approaches, probabilistic assessment techniques can provide an inclusive and realistic measure of system stability status [14]. Novel probabilistic assessment techniques are desirable and are being established. Probabilistic transient stability (PTS) assessment has been recognized to be a fitting approach to analyze the effect of uncertain parameters on transient stability [15]–[21]. Additionally, the results from PTS analysis can be associated with risk assessment, which is imperative for system operators, as economic and technical reasons can result in the power system to operate near the stability limit [21]. Although, it has been long established that deterministic studies may not sufficiently characterize the full extent of system dynamic behavior, the probabilistic approach has not been extensively used in the past in power system studies, mainly due to lack of data, limitation of computational resources, limited commercial softwares for probabilistic

U. Shahzad is with the Department of Electrical and Computer Engineering, University of Nebraska-Lincoln, Lincoln, NE, 68588-0511 USA (email: umair.shahzad@huskers.unl.edu).

analysis, deterministic nature of standards enforced by regulatory authorities, such as North American Electric Reliability Corporation (NERC), and mixed response from power utilities and planners [17], [20]. However, in recent times, there has been some research in PTS. For instance, [22] presented an analytical approach to determine PTS for online applications. [23]-[24] used Monte-Carlo Simulation (MCS) approach to present a stochastic-based approach to assess the PTS index. [15] proposed the inclusion of probabilistic considerations in transient stability investigation of a multimachine practical power system. Some other relevant work can be found in [25]-[30]. A major drawback of these works is that these use conventional numerical and analytical methods, such as, MCS, time-domain simulation, EEAC, TEF, hybrid method, etc. to estimate the transient stability index. These approaches may be suitable for an offline study, however, for real-time online prediction, a faster method is required. Artificial intelligence (AI)-based approaches provide a good alternative to fulfil this vital objective.

Among various AI approaches, Machine learning (ML) is an upcoming approach for solving power system problems, including transient stability [31]-[33]. ML is generally classified into three categories: supervised, unsupervised and reinforcement [34]. In supervised learning, the goal is to learn a mapping relation between the inputs to outputs, based on a given a labeled set of input/output pairs. Unsupervised learning deals with the training of an algorithm is using unlabeled data so that the algorithm may group the data based on similarity or difference. In reinforcement learning (RL), there is an interaction of an agent with its environment and consequently, the agent adapts its course, based on the reward because of its actions. The focus of this research work is Supervised Machine Learning (SML). Although, SML has various types [35], such as Artificial Neural Network (ANN), Decision Tree (DT), Random Forest, Support Vector Machine (SVM), etc., this work focuses on SVM-based SML for prediction of PTS.

In recent years, application of ML algorithms, such as ANN, to power system is an area of rising interest; the chief reason being the ability of ANN to process and learn intricate nonlinear relations [36]. Although, ANN is the most commonly used ML method for transient stability classification, it generally requires an extensive training process and an intricate design procedure. Moreover, ANN usually performs well for interpolation but not so well for extrapolation, which reduces its generalization ability. They are more susceptible to becoming trapped in a local minimum. Although, majority of ML algorithms can overfit if there is a dearth of training samples, but ANNs can also overfit if training goes on for a very long duration [37]. Due to these downsides, it becomes essential to develop a more efficient classifier for transient stability status prediction. SVM do not suffer from these drawbacks and has the following advantages, over ANN [38]: (1) less number of tuning parameters, (2) less susceptibility to overfitting, and (3) the complexity is depended on number of support vectors (SVs) rather than dimensionality of transformed input space.

Support vector machine (SVM) is an evolving ML approach that incapacitates some of the drawbacks of ANN. A SVM essentially is a SML algorithm that can use given data to solve certain problems by trying to convert them into linearly separable problems. Recently, SVM has been applied to power system transient stability classification problem. A SVM-based transient stability classifier was trained in [39] and its performance was compared with a multilayer perceptron (MLP) classifier. Reference [37] devised a multiclass SVM classifier for static and transient stability assessment and classification. Reference [40] suggested a SVM classifier to predict the transient stability status using voltage variation trajectory templates. Reference [38] trained a binary SVM classifier, with combinatorial trajectories inputs, to predict the transient stability status. Reference [41] employed the SVM to rank the synchronous generators based on transient stability severity and classify them into vulnerable and nonvulnerable machines. Reference [42] proposed two SVMs, using Gaussian kernels, for classifying the post-fault stability status of the system. Reference [43] presented a SVM-based approach for transient stability detection, using post-disturbance signals, from the optimally located distributed generations. Some other relevant work dealing with SVM-based transient stability classification can be found in [44]-[51]. Based on the detailed literature review and to the best of author's knowledge, there exists no research work on PTS which uses SVM-based SML approach, considering the uncertainties of load, faulted line, fault type, fault location (on the line), and FCT. Moreover, [52] specifically mentions the potential of SVM for online DSA, and [53]-[57] strongly indicate that ML is a promising and upcoming approach for online DSA. Thus, the main contribution of this paper is to predict PTS status using an SVM-based SML approach.

The rest of the paper is organized as follows. Section II describes various probabilistic factors associated with transient stability assessment. Section III discusses the PTS index used in this paper. Section IV provides a brief overview of SVM and its application to PTS classification problem. Section V discusses the procedure for the proposed approach. Section VI and VII deals with the description of case study, and associated results and discussion, respectively. Section VIII describes sensitivity analysis, with respect to some important parameters/functions. Finally, Section IX concludes the paper, with a suggested direction for future research.

## II. Probabilistic Factors in Power System Transient Stability

There are various factors which are involved in PTS assessment of power systems, such as fault type, fault location, load, and FCT. Suitable probability density functions (PDFs) are used to model these factors. The modeling approaches are described below [58]. Normally, shunt faults, such as three-phase (LLL), double-line-to-ground (LLG), line-to-line (LL) and single-line-to-ground (LG) short circuits, are considered for evaluating PTS. A probability mass functions (PMF) is normally used to model the fault type. Based on past system statistics, a usual practice is to select the probability of LLL, LL, LLG, and LG short circuits, as 0.05, 0.1, 0.15 and 0.7 respectively [22]. This paper adopts the same practice. The probability distribution of fault location on a transmission line is usually assumed to be uniform. This means that the fault can occur with equal probability at any line of the test system and at any point along the line [21]. This paper uses the same approach. The procedure of fault clearing constitutes of three stages: fault detection, relay operation and breaker operation. If the primary protection and breakers are 100% reliable, the clearing time is the only uncertain factor. A normal (Gaussian) PDF is generally used to model this time [21]. In this paper, fault is applied at 1 $s$ and it is cleared, after a mean time of 0.9 $s$ and standard deviation of 0.1 $s$ (based on the normal PDF). A normal PDF was used to represent the uncertainty of loads. Let $f(X)$ denote the PDF for individual bus loads, i.e.,

$$f(X) = \frac{1}{\sqrt{2\pi\sigma^2}} e^{\frac{-(X-\mu)^2}{2\sigma^2}} \qquad (1)$$

where $\mu$ and $\sigma$ denotes the mean and standard deviation of the forecasted peak load, respectively.

### III. QUANTIFICATION INDEX FOR PROBABILISTIC TRANSIENT STABILITY

The Transient Stability Index (*TSI*) was used to quantify the transient stability of a system consisting of synchronous machines [59]. This index is based on the maximum rotor angle separation between any two synchronous machines, after the fault has occurred. Mathematically, it is given by

$$TSI_i = \frac{360 - \delta_{\max i}}{360 + \delta_{\max i}}, \text{ where } -1 < TSI_i < 1 \qquad (2)$$

where $\delta_{\max i}$ is the post-fault maximum rotor angle separation (in degrees) between any two synchronous machines in the system at the same time (for a fault on $i^{th}$ line). A negative *TSI* value indicates that the power system is transiently unstable. This is a global index for a swift indication of the transient stability status of the system (for a fault on any line, at any point, for any FCT and for any load). Therefore, this index is used in this paper to quantify the PTS status. Let $S_i$ represent the PTS status indicator for $i^{th}$ iteration of MCS. Mathematically,

$$S_i = \begin{cases} 1, & \text{if } TSI_i < 0 \\ 0, & \text{if } TSI_i \geq 0 \end{cases} \qquad (3)$$

Therefore, if the system is transiently stable, for $i^{th}$ Monte-Carlo (MC) sample, value of $S_i$ will be 0; otherwise, it will be 1. This information will be used for training the SVM model.

### IV. SVM: BRIEF OVERVIEW AND APPLICATION TO PROBABILISTIC TRANSIENT STABILITY PREDICTION

Support vector machine (SVM), which is also known as maximum margin classifier, is a type of SML, that can be used both in classification and regression problems. It was first introduced by Vapnik [60]-[61] and was elaborated by Schölkopf et al. [62]. SVM classifiers depend on training points, which lie on the boundary of separation between different classes, where the evaluation of transient stability is important. A decent theoretical progress of the SVM, due to its basics built on the Statistical Learning Theory (SLT) [60], made it possible to develop fast training methods, even with large training sets and high input dimensions [63]-[65]. This useful characteristic can be applied to tackle the issue of high input dimension and large training datasets in the PTS problem. The basic implementation of an SVM, commonly known as a hard margin SVM, requires the binary classification problem to be linearly separable. This is frequently not the case in practical problems, and therefore, SVM provides a kernel trick to resolve this issue. The forte of the SVM algorithm is based on the use of this kernel trick to transform the input space into a higher dimensional feature space. This permits to define a decision boundary that linearly separates the classes. The SVM algorithm attempts to determine that decision boundary or hyperplane with the highest distance from each class [38], [66]. The hyperplane can be mathematically defined as [39]

$$(w^T x) + b = 0 \qquad (4)$$

where $w$ is the weight vector ($w^T$ is its transpose), $x$ is the sample feature vector, and $b$ is a bias value. The samples that assist the algorithm to define the optimal hyperplane are those that lie closest to it, and they are known as SVs. The kernel function plays a significant role in SVM classification [67]. The kernel function is applied on each data instance to map the original non-linear data points into a higher-dimensional space in which they become linearly separable. An SVM classifier minimizes the generalization error by optimizing the relation between the number of training errors and the so-called Vapnik-Chervonenkis (VC) dimension. This is attained using the approach of structural risk minimization (SRM) which states that the classification error expectation of unseen data is bounded by the sum of a training error rate and a term that depends on the VC dimension [39]. Compared to empirical risk minimization (ERM)-based formulation (which is used by most ML algorithms, including ANN), the SRM-based formulation allows the SVM to prevent overfitting problems, by defining an upper bound, on the expected risk. A formal theoretical bound exists for the generalization ability of an SVM, which depends on the number of training errors ($t$), the size of the training set ($N$), the VC dimension associated to the resulting classifier ($h$), and a chosen confidence measure for the bound itself ($\eta$) [39], [61], [68]:

$$R < \frac{t}{N} + \sqrt{\frac{h(\ln(\frac{2N}{h})+1) - \ln(\frac{\eta}{4})}{N}} \qquad (5)$$

The risk (or classification error expectation) $R$ represents the classification error expectation over all the population of input/output pairs, even though the population is only partially known. This risk is a measure of the actual generalization error and does not require prior knowledge of the probability distribution of the data. SLT derives inequality (5) to mean that the generalization ability of an SVM is measured by an upper limit of the actual error given by the right-hand side of (5), and this upper limit is valid with a probability of $1-\eta$ ($0<\eta<1$). As $h$ increases, the first summand of the upper bound (5) decreases and the second summand increases, such that there is a balanced compromise between the two terms (complexity and training error), respectively [39]. The SVMs used for binary classification problems are based on linear hyperplanes to separate the data, as shown in Fig. 1. The hyperplane (represented by dotted line in Fig. 1) is determined by an orthogonal vector *w* and a bias *b*, which identify the points that satisfy $(w^T x) + b = 0$. By determining a hyperplane which maximizes the margin of separation, denoted by $\rho$, it is instinctively anticipated that the classifier will have an improved generalization ability. The hyperplane having the largest margin on the training set can be completely determined by the points that lie closest to the hyperplane. Two such points are $x_1$ and $x_2$ as shown in in Fig. l (b), and they are known as SVs because the hyperplane (i.e., the classifier) is completely dependent on these vectors. Consequently, in their simplest form, SVMs learn linear decision rules as

$$f(x) = sign(w^T x + b) \quad (6)$$

so that (*w, b*) are determined as to correctly classify the training examples and to maximize $\rho$. For linearly separable data, as shown in Fig. 1, a linear classifier can be found such that the first summand of bound (5) is zero.

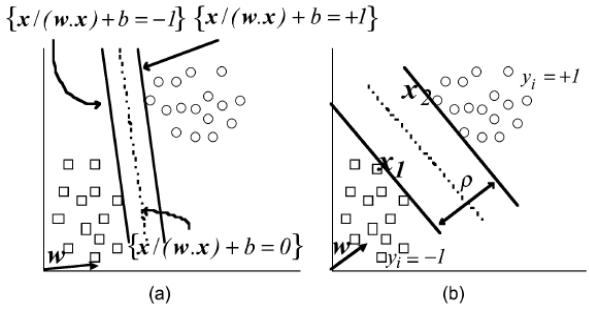

Fig. 1. SVM (maximum margin) classifier.

It is always possible to scale *w* and *b* such that

$$w^T x + b = \pm 1 \quad (7)$$

for the SVs, with

$$w^T x + b > +1 \text{ and } w^T x + b < -1 \quad (8)$$

for non-SVs.

Using the SVs $x_1$ and $x_2$ of Fig. 1 and (7), the margin $\rho$ can be calculated as

$$\rho = \frac{w^T}{||w||}(x_2 - x_1) = \frac{2}{||w||} \quad (9)$$

where $||w||$ is the Euclidean Norm of *w*. For linearly separable data, the VC dimension of SVM classifiers can be evaluated as

$$h < \min\left\{n, \frac{4D^2}{\rho^2}\right\} + 1 = \min\left\{n, D^2 ||w^2||\right\} + 1 \quad (10)$$

where *n* is the dimension of the training vectors and *D* is the minimum radius of a ball which contains the training points.

Thus, the risk (5) can be reduced by lessening the complexity of the SVM, that is, by increasing the margin of separation $\rho$, which is equivalent to reducing $||w||$. In practice, as the problems are not probable to be detachable by a linear classifier, thus, the linear SVM can be extended to a nonlinear version by mapping the training data to an expanded feature space using a nonlinear transformation:

$$\Phi(x) = (\phi_1(x), ......, \phi_m(x)) \in R^m \quad (11)$$

where *m > n*. Then, the maximum margin classifier of the data for the new space can be determined. With this method, the data points which are non-separable in the original space may become separable in the expanded feature space. The next step is to approximate the SVM by minimizing (i.e., maximizing $\rho$)

$$V(w) = \frac{1}{2} w^T . w \quad (12)$$

subject to the constraint that all training patterns are correctly classified, i.e.,

$$y_i \cdot \{w^T \cdot \Phi(x_i) + b\} \geq 1, \quad i = 1, ..., N \quad (13)$$

Though, contingent on the kind of nonlinear mapping (11), the samples of training data may not be linearly separable. In this case, it is not possible to find a linear classifier that satisfies all the conditions given by (12). Thus, instead of (12), a new cost function is optimized, i.e.,

$$\min V(w, \varepsilon) = \frac{1}{2} w^T . w + C \sum_{i=1}^{N} \varepsilon_i$$
$$s.t. \ y_i \cdot \{w^T \cdot \Phi(x_i) + b\} \geq 1 - \varepsilon_i \text{ for } i = 1, ..., N \quad (14)$$
$$\varepsilon_i \geq 0 \qquad \text{for } i = 1, ..., N$$

where *N* non-negative slack variables $\varepsilon_i$ are introduced to allow training errors (i.e., training patterns for which $y_i \cdot \{w^T \cdot \Phi(x_i) + b\} \geq 1 - \varepsilon_i$ and $\varepsilon_i > 1$) and allow for some misclassification. By minimizing the first summand of (14), the complexity of the SVM is reduced, and by minimizing the second summand of (14), the number of training errors is decreased. *C* is a positive penalty factor (also known as regularization factor or soft margin parameter) which decides the tradeoff between the two terms. In case it is small, the separating hyperplane is more focused on maximizing the margin (at the expense of larger classification mistakes), while the number of misclassified points is minimized for larger *C* values (at the expense of keeping the margin small). The minimization of the cost function (14) leads to a quadratic optimization problem with a unique solution. The nonlinear mapping (11) is indirectly obtained by the kernel functions, which correspond to inner products of data vectors in the expanded feature space $K(a, b) = \Phi(a)^T \cdot \Phi(b), \ a, b \in R^n$ [39], [46], [68]. Common kernel functions include the linear, polynomial, sigmoid and Gaussian radial basis function (RBF). The Gaussian RBF kernel generally is preferred over others because it has the ability of mapping samples nonlinearly into a higher dimensional space, and therefore, unlike linear kernel, it can tackle the scenario when the relationship between class labels and attributes is nonlinear. Although, sigmoid kernel performs like a Gaussian RBF kernel for certain parameters, but there are some parameters for which the sigmoid kernel is not the dot product of two vectors, thus, it is invalid. Moreover, as compared to polynomial kernel, it has few hyperparameters (parameters whose values are used to control the learning process) [61]. Thus, this work uses a Gaussian RBF kernel, which is mathematically given by,

$$K(a,b) = e^{-(\gamma\|a-b\|^2)}, \gamma > 0, \gamma = \frac{1}{2\sigma^2} \quad (15)$$

where $\gamma$ denotes the kernel parameter of the SVM classifier and $\sigma$ is the width of the Gaussian function.

The hyperparameters $C$ and $\gamma$ impact how sparse and easily separable the training data are in the expanded feature space. Subsequently, these parameters decide the complexity and training error rate of the resulting SVM classifier. These parameters must be optimized for achieving the best performance for the SVM classifier. The block diagram for the proposed SVM framework is shown in Fig. 2. The proposed SVM framework used has four inputs (system load, fault type, fault location and FCT), and one output (for $S_i$). Samples for training data were chosen using the MCS-based time domain simulation approach (described in Section V).

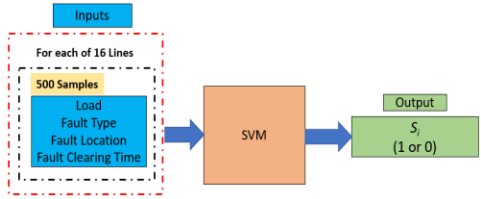

Fig. 2. Framework for the proposed SVM approach (input features and corresponding output).

For the PTS classification task, the first step was feature extraction, i.e., to select the most relevant input and output data for the SVM classification model. System load, fault type, fault location, and FCT were chosen as inputs, and transient stability status, $S_i$, was selected as the output (the binary variable to be classified as transiently stable or unstable). 500 samples were used for each line to train the SVM model, as shown in Table I. It must be mentioned that generally, there is no accepted rule of thumb to determine the number of samples for training the ML model; this typically depends on complexity of the problem, required performance level, and the ML algorithm used. As there are 16 lines in the system, thus, the total number of samples used for SVM model were 8000 ($500 \times 16$). Thus, the size of the input feature matrix was $8000 \times 4$. The Gaussian RBF kernel function was used for training the SVM as there is ample nonlinearity amongst the data presented to the SVM classifier. The hyperparameters $C$ and $\gamma$ were optimized using Bayesian optimization (other approaches such as Grid search or Random search may also be used). The data presented to SVM is randomly divided in two subsets: training subset and testing subset. The $K$-fold cross-validation approach is used to accomplish this as this prevents over fitting while training the data. In this approach, the entire data is divided into $K$ partitions of equal size. Training and testing are repeated, each time selecting a different partition for testing data, until all $K$ partitions are utilized for testing, i.e., every data point gets to be in a test set exactly once and gets to be in a training set ($K$-1) times [38]. Eventually, the average of these errors is taken as the expected prediction error. This work used the value of $K$ as 5, i.e., in each fold, 20% data was used for testing and 80% for training.

## V. Procedure for The Proposed Approach

The methodology for the proposed approach is described in Fig. 3. The IEEE-14 bus system was used to test and validate the proposed approach. This system has 16 transmission lines. For each line, 500 random MC samples were generated (the symbol $i$ indicates the sample number for the MCS). It is assumed that pre-fault system topology (configuration) is fixed, i.e., there is no contingency before the fault occurrence. In the first step, the first line is selected. In the next step, MCS is initiated with 500 samples. In each sample, system load, fault type, fault location and FCT are randomly chosen (based on the respective defined PDFs, as described in Section II). The fault is created at time $t=1$ s. For each MC sample, time-domain stability simulation is run for 10 s to determine the outcome (transiently stable or unstable). This is determined based on the value of $S_i$, as described in Section III. These steps are repeated and MCS is performed (for 500 samples) for all the remaining lines. When the MCS is run for all the 16 lines in the network, the resulting data obtained is used as training data for the SVM classification model.

A summarized workflow of SML application for online PTS prediction is shown in Fig. 4. As illustrated, the first step deals with the offline mode. In this mode, time-domain simulations are conducted, considering the uncertainties of input variables in the form of PDFs (generally obtained from past historical observations). In the next step, these distributions are sampled to gather enough training data. For each sample, the PTS status is measured by a binary variable, say, $x$, which can take two labels (say, 1 for transiently unstable, and 0 for transiently stable). Therefore, the final training data consists of the PTS status labels and the corresponding input operating conditions. In the next step, this offline-based database is used for online PTS prediction. The SML model 'learns' the stability rules and consequently, can be used to predict the PTS status for current operating point.

## VI. Case Study

The IEEE 14-bus test transmission system was used to conduct the required simulations. The numerical data and parameters were taken from [69]. The single line diagram is shown in Fig. 5. It should be highlighted that the proposed methodology is applicable to any test system. As mentioned before, a normal PDF is used to define the uncertainty in system loads. The active power of each load was assigned a mean equal to the original load active power value, as given in test system data in [69], and a standard deviation equal to 10% of the mean value. All time-domain simulations are RMS simulations and are performed using DIgSILENT PowerFactory software [70]. For SML application, Classification Learner tool of MATLAB was used [71].

TABLE I
SELECTED FEATURES FOR EACH LINE

| Feature Name | Number of Samples |
|---|---|
| System load | 500 |
| Fault type | 500 |
| Fault location | 500 |
| FCT | 500 |

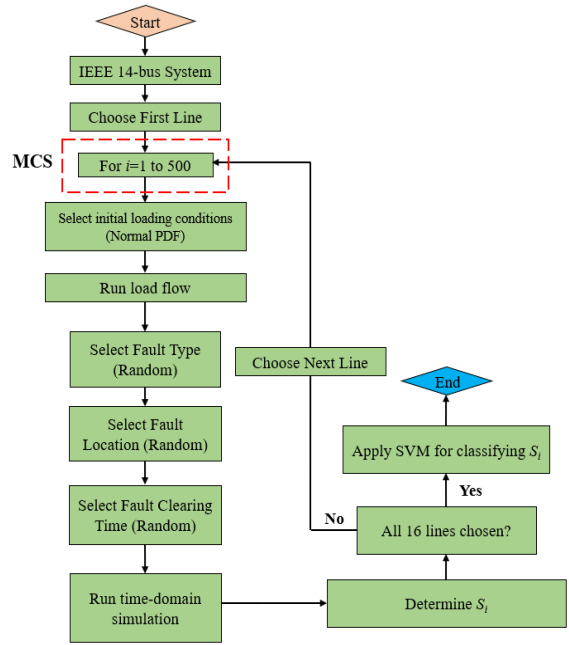

Fig. 3. Flowchart for the proposed SVM approach.

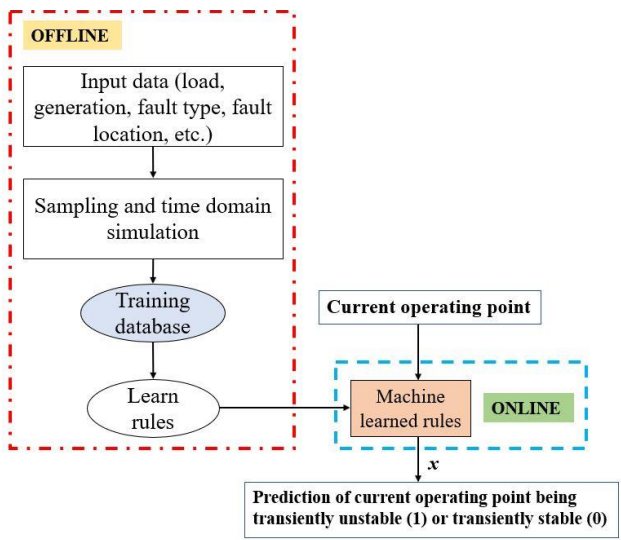

Fig. 4. Proposed SML approach for online PTS prediction.

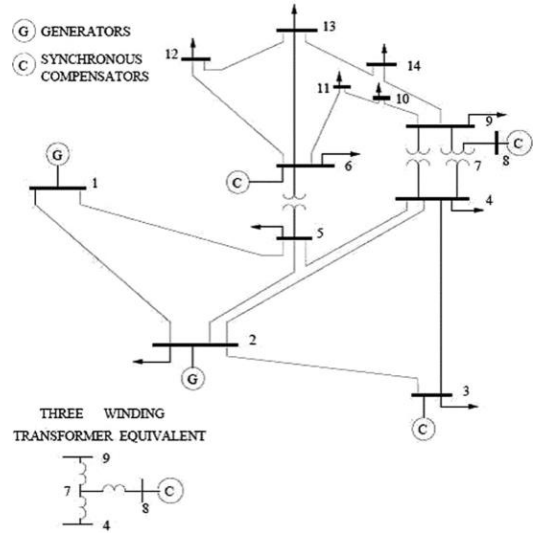

Fig. 5. IEEE 14-bus test system.

## VII. RESULTS AND DISCUSSION

It is assumed that required accuracy of the classifier must be more than 95%. To quantify the performance of the trained SVM classifier, the confusion matrix was used. This matrix is a graphical representation of the number of samples predicted correctly and incorrectly. The confusion matrix obtained for PTS classification is shown in Fig. 6 (1 and 0 represents transiently unstable and transiently stable class, respectively). Classification accuracy ($CA$) is a commonly used classification performance metric [72]. It is calculated as the proportion of correct predictions from the total number of the data points. The ideal value of $CA$ is 1, whereas the worst is 0. It is mathematically defined as

$$CA = \frac{TP+TN}{TP+TN+FP+FN} = \frac{TP+TN}{N} \quad (16)$$

where $TP$ (true positives), $TN$ (true negatives) denotes the correctly predicted data, and $FP$ (false positives), $FN$ (false negatives) denotes the incorrectly predicted data. $N$ denotes total data points (which are 8000 in this paper). Similarly, classification error ($CE$) represents the number of incorrect predictions from the total number of the data points. The closer it is to zero, the better. Mathematically,

$$CE = \frac{FP+FN}{TP+TN+FP+FN} = \frac{FP+FN}{N} \quad (17)$$

A receiver operating characteristic (ROC) curve is a graphical plot that establishes the diagnostic ability of a binary ML classifier [73]. In this plot, the true positive rate (sensitivity) is plotted against the false positive rate (1-specificity). Sensitivity is a measure of actual positives which are correctly identified, whereas, specificity is the proportion of truly negative cases that were classified as negative [74]. A classification SVM model with perfect discrimination has a ROC plot that passes through the upper left corner (100% sensitivity, 100% specificity), i.e., its area under curve ($AUC$) is equal to 1. The closer the $AUC$ is to 1, the greater the classification accuracy. The confusion matrix and the ROC curve (for testing data), for the classification of $S_i$, are shown in Fig. 6 and Fig. 7, respectively. From Fig. 6, it is evident that $CA$ for the confusion matrix is very high, i.e., approximately 97% (59.46% + 37.23%). Moreover, as evident from Fig. 7, the ROC curve is very accurate ($AUC > 0.99$). The values of various classification metrics are summarized in Table II. As evident, values for $CA$ and $AUC$ are in the desired high accuracy range (>0.95), and $CE$ is quite small (0.033). Once trained, the SVM classifier can be directly used to classify $S_i$. The training time for the SVM classifier was only 0.03 $s$. Thus, it can be inferred that the trained SVM algorithm can rapidly classify the PTS status, $S_i$, with a high accuracy (≈97%). This makes it suitable for an online application and therefore, can drastically help power system operators in the control center for decision-making tasks.

Thus, to sum up, the proposed SVM approach can be used to predict the PTS status, incorporating various uncertain factors (system load, faulted line, fault type, fault location, and FCT), with a superior accuracy. This approach has an edge over the conventional approaches, as it is computationally efficient, as well as, fairly accurate. It is strongly believed that the proposed approach can drastically contribute to progressing the prevailing methods for online DSA.

It must be mentioned that the proposed ML algorithm is system-specific and, although, it performed quite well for the IEEE 14-bus system, it is not assured that it will perform the same for other systems. Therefore, ample testing and validation of the proposed approach must be conducted on other standard test systems, before reaching a generic conclusion on the performance of ML algorithm. Additional generic limitations exist for ML-based approach, for instance, the training database and ML model must be updated when the PDFs of the input random variables, and the network topology varies over time, and consequently, the number of transient stability simulations required for training may be greater than that estimated for a fixed topology. An additional limitation regarding SVM is that it is sensitive to noise (target classes overlap) and outliers (target classes deviate significantly from the rest of the classes), and consequently, does not give a good performance. Moreover, choosing the optimal kernel function is not straightforward and may require several optimization simulations [75]. Also, the best ML approach may change depending on the application [55]. An avid reader can refer to [76] for further details.

## VIII. SENSITIVITY ANALYSIS

As mentioned before, the value of $K$ used in this work was 5. To verify that it is indeed the best value, a sensitivity analysis was performed. The SVM classifier was trained for various values of $K$, and the corresponding $CA$ values were determined. The results obtained are shown in Fig. 8. As evident, increasing $K$ beyond 5 does not alter the $CA$. Hence, $K=5$ is a good choice for $K$-fold cross-validation, for this work. This also validates the fact that $K=5$ and $K=10$ are generally the most commonly used values for a $K$-fold cross-validation procedure [77]. Moreover, for $K=5$, the values of $CA$, for different kernel functions, are shown in Table III. As evident, Gaussian RBF kernel has the highest accuracy. This also validates the reason of Gaussian RBF kernel being the most commonly used kernel function for SVM classification [61].

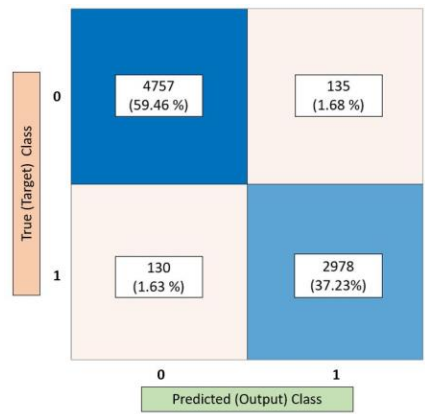

Fig. 6. Confusion matrix for transient stability classification performance assessment.

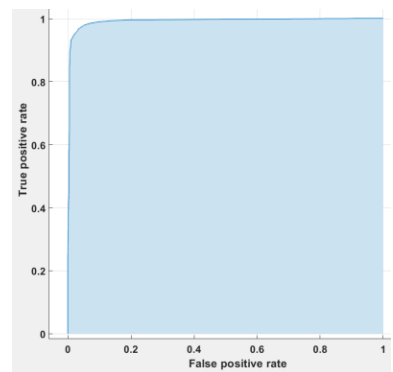

Fig. 7. ROC curve for transient stability classification performance assessment.

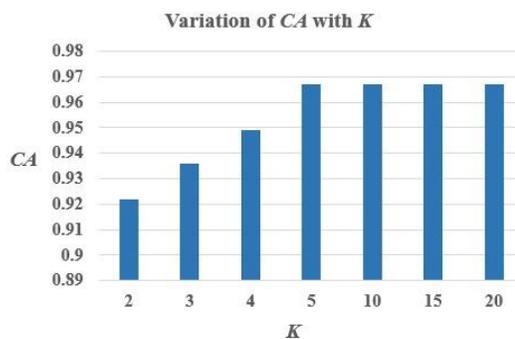

Fig. 8. Variation of $CA$ with $K$.

## IX. CONCLUSION AND FUTURE WORK

The paper highlighted the need to consider a faster method for PTS assessment and hence, proposed a binary SVM

approach for predicting PTS status. In addition to uncertain system load conditions, various uncertain factors such as faulted line, fault type, fault location and FCT were considered. Time-domain simulations were used to gather the data required for training the SVM model. The *TSI* was used as the indicator for the PTS status. The proposed method was applied to the IEEE 14-bus system, and promising results were obtained, indicating the significance of SVM in power system PTS assessment. The results indicated that the proposed approach predicted the PTS status with an excellent accuracy, in a computationally efficient manner. This indicates the potential of SVM for online DSA, especially for large-scale power systems.

As a future work, ensemble learning, incorporating multiple learning methods, can be applied for prediction of PTS. Moreover, incorporating data from renewable energy generation sources (such as wind and solar) in the SVM training model can prove to be very useful in online DSA procedure.

TABLE II
SVM PERFORMANCE ASSESSMENT USING VARIOUS CLASSIFICATION METRICS

| Classification Metric | Value |
|---|---|
| CA | 0.967 |
| CE | 0.033 |
| AUC | 0.991 |

TABLE III
VARIATION OF *CA* FOR DIFFERENT KERNEL FUNCTIONS

| Kernel Function | CA |
|---|---|
| Linear | 0.872 |
| Polynomial (order 2) | 0.916 |
| Polynomial (order 3) | 0.829 |
| Gaussian RBF | 0.967 |

REFERENCES

[1] M. P. Bhavaraju, R. Billinton, R. E. Brown, J. Endrenyi, W. Li, A. P. Meliopoulos, and C. Singh, "IEEE Tutorial on electric delivery system reliability evaluation," in *Proc. IEEE Power Energy Soc. Gen. Meet.*, 2005, pp. 68-78.
[2] P. Kundur, J. Paserba, V. Ajjarapu, G. Andersson, A. Bose, C. Canizares, N. Hatziargyriou, D. Hill, A. Stankovic, C. Taylor, T. V. Cutsem, and V. Vittal, "Definition and classification of power system stability IEEE/CIGRE joint task force on stability terms and definitions," *IEEE Trans. Power Syst.*, vol. 19, no. 3, pp. 1387-1401, Aug. 2004.
[3] A. M. A. Haidar, M. W. Mustafa, F. A. F. Ibrahim, and I. A. Ahmed, "Transient stability evaluation of electrical power system using generalized regression neural networks," *Appl. Soft Comp.*, vol. 1, no. 4, pp. 3558-3570, Jun. 2011.
[4] A. L. Bettiol, A. Souza, J. L. Todesco, and J. R. Tesch, "Estimation of critical clearing times using neural networks," in *Proc. IEEE Bologna Pow. Tech Conf.*, 2003, pp. 1-6.
[5] A. Karami, "Power system transient stability margin estimation using neural networks," *Int. J. Electr. Power Energy Syst.*, vol. 33, no. 4, pp. 983-991, May 2011.
[6] M. A. Pai, *Energy Function Analysis for Power System Stability*. Boston, MA, USA: Kluwer, 1989.
[7] A. A. Fouad and V. Vittal, *Power System Transient Stability Analysis Using the Transient Energy Function Method*. Upper Saddle River, NJ, USA: Prentice-Hall, 1992.
[8] Y. Xue, T. V. Cutsem, and M. Ribbens-Pavella, "Extended equal area criterion justifications, generalizations, applications," *IEEE Trans. Power Syst.*, vol. 4, no. 1, pp. 44-52, Feb. 1989.
[9] L. D. Colvara, "Stability analysis of power systems described with detailed models by automatic method," *Int. J. Electr. Power Energy Syst.*, vol. 31, no. 4, pp. 139-145, May 2009.
[10] H. Sawhney and B. Jeyasurya, "A feed-forward artificial neural network with enhanced feature selection for power system transient stability assessment," *Elect. Power Syst. Res.,* vol. 76, no. 12, pp. 1047-1054, Aug. 2006.
[11] J. G. Slootweg and W. L. Kling, "Impacts of distributed generation on power system transient stability," in *Proc. IEEE Power Energy Soc. Gen. Meet.*, 2002, pp. 862-867.
[12] A. M. Azmy and I. Erlich, "Impact of distributed generation on the stability of electrical power system," in *Proc. IEEE Power Energy Soc. Gen. Meet.*, 2005, pp. 1056-1063.
[13] A. Khosravi, M. Jazaeri, and S. A. Mousavi, "Transient stability evaluation of power systems with large amounts of distributed generation," in *Proc. Intl. Uni. Power Eng. Conf.*, 2010, pp. 1-5.
[14] X. Zhao and J. Zhou, "Probabilistic transient stability assessment based on distributed DSA computation tool," in *Proc. PMAPS.,* 2010, pp. 685-690.
[15] R. Billinton and P. R. S. Kuruganty, "Probabilistic assessment of transient stability in a practical multimachine system," *IEEE Trans. Power App. Syst.*, vol. PAS-100, no. 7, pp. 3634–3641, Jul. 1981.
[16] Y.-Y. Hsu and C. Chung-Liang, "Probabilistic transient stability studies using the conditional probability approach," *IEEE Trans. Power Syst.*, vol. 3, no. 4, pp. 1565–1572, Nov. 1988.
[17] E. Vaahedi, W. Li, T. Chia, and H. Dommel, "Large scale probabilistic transient stability assessment using B.C. Hydro's on-line tool," *IEEE Trans. Power Syst.*, vol. 15, no. 2, pp. 661–667, May 2000.
[18] P. M. Anderson and A. Bose, "A probabilistic approach to power system stability analysis," *IEEE Trans. Power App. Syst.*, vol. PAS-102, no. 8, pp. 2430–2439, Aug. 1983.
[19] K. J. Timko, A. Bose, and P. M. Anderson, "Monte Carlo simulation of power system stability," *IEEE Trans. Power App. Syst.*, vol. PAS-102, no. 10, pp. 3453–3459, Oct. 1983.
[20] W. Li, "Probabilistic transient stability assessment," in *Risk Assessment of Power Systems: Models, Methods, and Applications*. New York, NY, USA: Wiley-IEEE Press, 2014.
[21] P. N. Papadopoulos and J. V. Milanović, "Probabilistic framework for transient stability assessment of power systems with high penetration of renewable generation," *IEEE Trans. Power Syst.,* vol. 32, no. 4, pp. 3078-3088, Jul. 2017.
[22] M. Abapour and M. Haghifam, "Probabilistic transient stability assessment for on-line applications," *Int. J. Electr. Power Energy Syst.*, vol. 42, no. 1, pp. 627-634, Nov. 2012.
[23] J. Fang, W. Yao, J. Wen, S. Cheng, Y. Tang, and Z. Cheng, "Probabilistic assessment of power system transient stability incorporating SMES," *Phy. C: Supercond.*, vol. 484, pp. 276-281, Jan. 2013.
[24] S. O. Faried, R. Billinton, and S. Aboreshaid, "Probabilistic evaluation of transient stability of a power system incorporating wind farms," *IET Ren. Power Gen.*, vol. 4, no. 4, pp. 299-307, Jul. 2010.
[25] S. O. Faried, R. Billinton, and S. Aboreshaid, "Probabilistic evaluation of transient stability of a wind farm," *IEEE Trans. Energy Convers.*, vol. 24, no. 3, pp. 733-739, Sep. 2009.
[26] R. Billinton and P. R. S. Kuruganty, "Probabilistic considerations in transient stability assessment," *Canad. Electr. Eng. J.*, vol. 4, no. 2, pp. 26-30, Apr. 1979.
[27] L. Shi, S. Sun, L. Yao, Y. Ni, and M. Bazargan, "Effects of wind generation intermittency and volatility on power system transient stability," *IET Ren. Power Gen.*, vol. 8, no. 5, pp. 509-521, Jul. 2014.
[28] D. Han, J. Ma, A. Xue, T. Lin, and G. Zhang, "The uncertainty and its influence of wind generated power on power system transient stability under different penetration," in *Proc. Int. Conf. Power Syst. Tech.*, 2014, pp. 675-680.

[29] M. Al-Sarray, H. Mhiesan, M. Saadeh ,and R. McCann, "A probabilistic approach for transient stability analysis of power systems with solar photovoltaic energy sources," in *Proc. GreenTech*, 2016, pp. 159-163.

[30] P. Ju, H. Li, C. Gan, Y. Liu, Y. Yu, and Y. Liu, "Analytical assessment for transient stability under stochastic continuous disturbances," *IEEE Trans. Power App. Syst.*, vol. 33, no. 2, pp. 2004-2014, Mar. 2018.

[31] D. J. Sobajic and Y. Pao, "Artificial neural-net based dynamic security assessment for electric power systems," *IEEE Trans. Power App. Syst.*, vol. 4, no. 1, pp. 220-228, Feb. 1989.

[32] M. Djukanovic, D. J. Sobajic, and Y. Pao, "Neural-net based unstable machine identification using individual energy functions," *Int. J. Electr. Power Energy Syst.*, vol. 13, no. 5, pp. 255-262, Oct. 1991.

[33] Y. Pao and D. J. Sobajic, "Combined use of unsupervised and supervised learning for dynamic security assessment," *IEEE Trans. Power App. Syst.*, vol. 7, no. 2, pp. 878-884, May 1992.

[34] M. S. Ibrahim, W. Dong, and Q. Yang, "Machine learning driven smart electric power systems: Current trends and new perspectives," *Appl. Energy*, vol. 272, pp. 1-19. Aug. 2020.

[35] R. Saravanan and P. Sujatha, "A state of art techniques on machine learning algorithms: a perspective of supervised learning approaches in data classification," in *Proc. ICICCS*, 2018, pp. 945-949.

[36] T. S. Dillon, and D. Niebur, *Neural Networks Applications in Power Systems*. Leicestershire, UK: CRL Publishing, 1996.

[37] S. Kalyani and K. S. Swarup, "Classification and assessment of power system security using multiclass SVM," *IEEE Trans. Syst., Man, Cybern A., Syst., Humans*, vol. 41, no. 5, pp. 753-758, Sep. 2011.

[38] D. You, K. Wang, L. Ye, J. Wu, and R. Huang, "Transient stability assessment of power system using support vector machine with generator combinatorial trajectories inputs", *Int. J. Electr. Power Energy Syst.*, vol. 44, no. 1, pp. 318-325, Jan. 2013.

[39] L. S. Moulin, A. P. A. da Silva, M. A. El-Sharkawi, and R. J. Marks, "Support vector machines for transient stability analysis of large-scale power systems," *IEEE Trans. Power App. Syst.*, vol. 19, no. 2, pp. 818-825, May 2004.

[40] A. D. Rajapakse, F. Gomez, K. Nanayakkara, P. A. Crossley, and V. V. Terzija, "Rotor angle instability prediction using post-disturbance voltage trajectories," *IEEE Trans. Power App. Syst.,* vol. 25, no. 2, pp. 947-956, May 2010.

[41] B. P. Soni, A. Saxena, V. Gupta, and S. L. Surana, "Assessment of transient stability through coherent machine identification by using least-square support vector machine", *Mod. Simul. Eng.*, vol. 2018, pp. 1-18, May 2018.

[42] F. R. Gomez, A. D. Rajapakse, U. D. Annakkage, and I. T. Fernando, "Support vector machine-based algorithm for post-fault transient stability status prediction using synchronized measurements," *IEEE Trans. Power App. Syst.*, vol. 26, no. 3, pp. 1474-1483, Aug. 2011.

[43] P. Pavani and S. N. Singh, "Support vector machine based transient stability identification in distribution system with distributed generation," *Electr. Power Comp. Syst.,* vol. 44, no. 1, pp. 60-71, Nov. 2015.

[44] A. E. Gavoyiannis, D. G. Vogiatzis, D. R. Georgiadis, and N. D. Hatziargyriou, "Combined support vector classifiers using fuzzy clustering for dynamic security assessment," in *Proc. IEEE Power Energy Soc. Gen. Meet.*, 2001, pp. 1281-1286.

[45] S. Ye, X. Li, X. Wang, and Q. Qian, "Power system transient stability assessment based on adaboost and support vector machines," in *Proc. Asia-Pacific Power Ener. Eng. Conf.*, 2012, pp. 1-4.

[46] L. S. Moulin, A. P. A. da Silva, M . A. El-Sharkawi, and R. J. Marks, "Neural networks and support vector machines applied to power systems transient stability analysis," *Int. J. Eng. Intel. Syst. Elect. Eng. Comm.*, vol. 9, no. 4, pp. 205-211, Nov. 2001.

[47] N. I. A. Wahab, A. Mohamed, and A. Hussain, "Transient stability assessment of a power system using PNN and LS-SVM methods," *J. Appl. Sci.*, vol. 7, no. 21, pp. 3208-3216, 2007.

[48] B. D. A. Selvi and N. Kamaraj, "Investigation of power system transient stability using clustering based support vector machines and preventive control by rescheduling generators," in *Proc. ICTES*, 2007, pp. 137-142.

[49] N. G. Baltas, P. Mazidi, J. Ma, F. A. Fernandez, and P. Rodriguez, "A comparative analysis of decision trees, support vector machines and artificial neural networks for on-line transient stability assessment," in *Proc SEST*, 2018, pp. 1-6.

[50] M. Arefi and B. Chowdhury, "Ensemble adaptive neuro fuzzy support vector machine for prediction of transient stability," in *Proc. NAPS*, 2017, pp. 1-6.

[51] E. A. Frimpong, P. Y. Okyere, and J. Asumadu, "Prediction of transient stability status using Walsh-Hadamard transform and support vector machine," in *Proc. IEEE PES PowerAfrica*, 2017, pp. 301-306.

[52] B. D. A. Selvi and N. Kamaraj, "Support vector regression machine with enhanced feature selection for transient stability evaluation," in *Asia-Pacific Pow. and Energy Eng. Conf.*, 2009, pp. 1-5.

[53] A. Sabo and N. I. A. Wahab, "Rotor angle transient stability methodologies of power systems: a comparison," in *IEEE Student Conf. on Res. and Dev.*, 2019, pp. 1-6.

[54] W. Li and J. Zhou, "Probabilistic reliability assessment of power system operations," *Electr. Power Comp. Syst.*, vol. 36, no. 10, pp. 1102-1114, Sep. 2008.

[55] J. L. Cremer and G. Strbac, "A machine-learning based probabilistic perspective on dynamic security assessment," *Int. J. Electr. Power Energy Syst.*, vol. 128, pp. 1-15, Jun. 2021.

[56] H. Yuan, J. Tan, and Y. Zhang, "Machine learning-based security assessment and control for bulk electric system operation," NREL, Feb. 2020. [Online]. Available: https://www.nrel.gov/docs/fy21osti/76089.pdf

[57] T. Zhang, M. Sun, J. L. Cremer, N. Zhang, G. Strbac, and C. Kang, "A confidence-aware machine learning framework for dynamic security assessment," *IEEE Trans. Power Syst.*, pp. 1-14, Feb. 2021.

[58] K. Jayashree and K. S. Swarup, "A distributed computing environment for probabilistic transient stability analysis," in *16th Nat. Power Syst. Conf.,* 2010, pp. 329-335.

[59] P. N. Papadopoulos and J. V. Milanović, "Impact of penetration of non-synchronous generators on power system dynamics," in *Proc. IEEE Eindhoven Power Tech.*, 2015, pp. 1-6.

[60] V. Vapnik, *Statistical Learning Theory*. New York, NY, USA: John Wiley & Sons, 1998.

[61] V. Vapnik, *The Nature of Statistical Learning Theory*. New York, NY, USA: Springer-Verlag, 1995.

[62] B. Schölkopf, C. Burges, and A. Smola, *Advances in Kernel Methods—Support Vector Learning*. Cambridge, MA, USA: MIT Press, 1999.

[63] E. Osuna, R. Freund, and F. Girosi, "Training support vector machines: au application to face detection," in *Proc. CVPR*, 1997, pp. 1-8.

[64] T. Joachims, "Making large-scale SVM learning practical," in *Advances in Kernel Methods - Support Vector Learning*. Cambridge, MA, USA: MIT Press, 1998.

[65] N. Cristianini and J. Shawe-Taylor, *An Introduction to Support Vector Machines and Other Kernel-Based Learning Methods*. Cambridge, UK: Cambridge University Press, 2000.

[66] G. N. Baltas, P. Mazidi, F. Fernandez, and P. Rodríguez, "Support vector machine and neural network applications in transient stability," in *Proc. ICRERA*, 2018, pp. 1010-1015.

[67] C. Savas and F. Dovis, "The impact of different kernel functions on the performance of scintillation detection based on support vector machines," *Sensors*, vol. 19, no. 23, pp. 1-16, Nov. 2019.

[68] L. S. Moulin, A. P. A. da Silva, M. A. El-Sharkawi, and R. J. Marks, "Support vector and multilayer perceptron neural networks applied to power systems transient stability analysis with input dimensionality reduction," in *Proc. IEEE Power Energy Soc. Gen. Meet.*, 2002, pp. 1308-1313.

[69] Power Systems Test Case Archive. [Online]. Available: http://www.ee.washington.edu/research/pstca/pf14/pg_tca14bus.htm

[70] DIgSILENT PowerFactory User Manual, DIgSILENT GmbH, 2018. [Online]. Available: https://www.digsilent.de/en/downloads.html

[71] Classification Learner. [Online]. Available: https://www.mathworks.com/help/stats/classificationlearner-app.html

[72] N. I. A. Wahab, A. Mohamed, and A. Hussain, "Fast transient stability assessment of large power system using probabilistic neural network with feature reduction techniques," *Exp. Syst. Appl.*, vol. 38, no. 9, pp. 11112-11119, Sep. 2011.

[73] Using the ROC curve to analyze a classification model. [Online]. Available: http://www.math.utah.edu/~gamez/files/ROC-Curves.pdf

[74] C. Sammut and G. I. Webb, *Encyclopedia of Machine Learning*. New York, NY, USA: Springer, 2011.
[75] S. Karamizadeh, S. M. Abdullah, M. Halimi, J. Shayan and M. J. Rajabi, "Advantage and drawback of support vector machine functionality," in *International Conf. on Comput., Commun., and Contr. Tech.*, 2014, pp. 63-65.
[76] L. Duchesne, E. Karangelos and L. Wehenkel, "Recent developments in machine learning for energy systems reliability management," *Proc. IEEE*, vol. 108, no. 9, pp. 1656-1676, Sep. 2020.
[77] J. Brownlee, A gentle introduction to k-fold cross-validation, 2018. [Online]. Available: https://machinelearningmastery.com/k-fold-cross-validation/

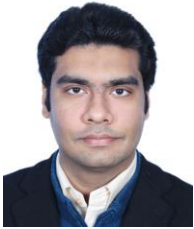
**Umair Shahzad** received a B.Sc. electrical engineering degree from the University of Engineering and Technology, Lahore, Pakistan, and a M.Sc. electrical engineering degree from The University of Nottingham, England, in 2010 and 2012, respectively. He is currently working toward a Ph.D. electrical engineering from the University of Nebraska-Lincoln, Lincoln, NE, USA. His research interests include power system analysis, power system security assessment, power system stability, machine learning, and probabilistic methods applied to power systems.